\documentclass[12pt]{article}
\usepackage{graphicx}
\begin{document}
\renewcommand{\thefootnote}{\fnsymbol{footnote}}
\begin{titlepage}
\begin{flushright}
OU-HET 474 \\
hep-th/0406069
\end{flushright}

\vspace{10mm}
\begin{center}
{\Large\bf Hamiltonian Analyses of New Set of Dual Actions of Bosonic $p$-Branes}
\vspace{25mm}

{\large
Yan-Gang Miao\footnote{e-mail address:
 miaoyg@nankai.edu.cn; miao@het.phys.sci.osaka-u.ac.jp}}\\
\vspace{10mm} $^{1)}${\em Department of Physics, Nankai
University, Tianjin 300071, \\
People's Republic of
China\footnote{Permanent address.}}

$^{2)}${\em Department of Physics, Osaka University,
Toyonaka, Osaka 560-0043, Japan}

\end{center}

\vspace{15mm}
\centerline{{\bf{Abstract}}}
\vspace{5mm}
We make Hamiltonian analyses of the new set of dual bosonic $p$-brane (including string) actions which contain high non-linearity. The difficulties exist in two basic steps of the Hamiltonian procedure, that is, in  calculating canonical momenta and in solving velocities in terms of momenta. The former difficulty can be overcome by an ADM reparametrization of induced metrics, while the latter may be circumvented by some modification to the usual Hamiltonian procedure. We also compare our results with that of the other set composed of known dual $p$-brane actions.


\end{titlepage}
\newpage
\renewcommand{\thefootnote}{\arabic{footnote}}
\setcounter{footnote}{0}
\setcounter{page}{2}

\section{Introduction}
Bosonic $p$-branes~\cite{s1} are extended objects that are embedded in
a higher dimensional spacetime, and their dynamics is governed in general by two sets of dual actions~\cite{s2} in which they are symbolized by ($S_{NG}$, $S_{P}$, $S^{{\rm I}}_{W}$, $S^{{\rm II}}_{W}$) and ($S_{NG}$, $A_{P}$, $A^{{\rm I}}_{W}$, $A^{{\rm II}}_{W}$). The first set consists of four known actions, while the second contains three new ones where $A_{P}$ is not Weyl-invariant but $A^{{\rm I}}_{W}$ and $A^{{\rm II}}_{W}$ are Weyl-invariant. The seven different formulations of actions are classically equivalent from the point of view that they lead to the same equations of motion. Each set is closed by dualization with respect to various field variables appeared in actions. The second, {\em i.e.} the new set is discovered by using the paraent or master action approach~\cite{s3} together with a special proposal of parent actions. (For details of derivation and discussion of duality, see ref.~\cite{s2}.)

In the first set of dual actions, the
Nambu-Goto action $S_{NG}$ is proportional to a $(p+1)$-dimensional
worldvolume by Nambu~\cite{s4} and Goto~\cite{s5} for a string ($p=1$).
$S_P$ is the $p$-brane action with an auxiliary worldvolume metric and
a cosmological term, where the subscript means $p$-branes, whose
formulation of a membrane ($p=2$) was proposed by Howe and Tucker~\cite{s6}
in the construction of a locally supersymmetric invariant model of
a spinning membrane. Quite noticeable is that $S_P$ is not Weyl-invariant
for the general cases, $p \geq 2$, but Weyl-invariant for the string theory,
{\em i.e.}, $p=1$. This string action was first given by Brink, Di Vecchia and
Howe and Deser and Zumino~\cite{s7,s8}, and its virtues in path integral quantization were emphasized by Polyakov~\cite{s9}.
For $p$-brane ($p \geq 2$) cases a first order form of Weyl-invariant actions\footnote{This action is not discussed in our paper because it cannot be classified into any of the two sets of dual actions. For details, see ref.~\cite{s2}.} was first established~\cite{s10} from the Nambu-Goto action through directly replacing the induced metric by a Wely-invariant combination of the worldvolume metric and the induced metric itself. $S^{{\rm I}}_{W}$ and $S^{{\rm II}}_{W}$ called in  ref.~\cite{s2} first and second Weyl-invariant actions respectively, possess a Weyl symmetry. They were constructed~\cite{s11,s12,s13,s1} partly because of the important role played by the Weyl symmetry in covariant quantization of strings.

In the new set of dual actions, except for the
Nambu-Goto action which appears in both sets, the other three actions are newly constructed~\cite{s2}. $A_{P}$ which corresponds to $S_{P}$ does not possess the Weyl invariance {\em even} for the string theory, while $A^{{\rm I}}_{W}$ and $A^{{\rm II}}_{W}$ called in ref.~\cite{s2} first and second {\em new} Weyl-invariant actions respectively, are thus proposed with the aim at recovering such a symmetry. In particular, the three new actions contain high non-linearity of derivatives of spacetime with respect to worldvolume parameters which originates from a special definition of contravariant components of induced metrics. They have the following forms:
\begin{equation}
A_{P}=-\frac{T}{2}\int d^{p+1}{\xi}\sqrt{-g}\left[-g_{ij}h^{ij}+(p+3)\right],
\end{equation}

\begin{equation}
A^{{\rm I}}_{W}=-\frac{T}{2}\int d^{p+1}{\xi}\sqrt{-g}
\left[-{\Phi}^{(p+3)/2}g_{ij}h^{ij}+(p+3){\Phi}^{(p+1)/2}\right],
\end{equation}
and
\begin{equation}
A^{{\rm II}}_{W}=-T\int d^{p+1}{\xi}\sqrt{-g}\left(\frac{1}{p+1}g_{ij}
h^{ij}\right)^{-(p+1)/2},
\end{equation}
where $T$ is the $p$-brane tension. We have utilized the same notation as that in ref.~\cite{s2} and, for the purpose of explicitness, rewrite it as follows:
\begin{equation}
{\eta}_{{\mu}{\nu}}={\rm diag}(-1,1,\cdots,1),
\end{equation}
is the flat metric of a $D$-dimensional Minkowski spacetime.
Greek indices (${\mu},{\nu},{\sigma},\cdots$) run over $0,1,\cdots,D-1$.
\begin{equation}
h_{ij}\equiv \frac{{\partial}X^{\mu}}{{\partial}{\xi}^{i}}
\frac{{\partial}X^{\nu}}{{\partial}{\xi}^{j}}{\eta}_{{\mu}{\nu}},
\end{equation}
is defined as the induced metric in the $(p+1)$-dimensional worldvolume
spanned by $p+1$ arbitrary parameters ${\xi}^i$. Latin indices
($i,j,k,\cdots$) take the values $0,1,\cdots,p$. Coordinates
$X^{\mu}({\xi}^i)$ are a $D$-component Lorentz vector field in the
spacetime and $D$ scalar fields in the worldvolume as well.
$g_{ij}$ stands for the auxiliary worldvolume metric, $g^{ij}$ its inverse, and $g\equiv {\rm det}(g_{ij})$. ${\Phi}({\xi}^i)$ is introduced as a scalar in both the spacetime and
worldvolume in order to keep eq.~(2) invariant under the Lorentz
transformation and reparametrization. Especially, note that $h^{ij}$ is not $g^{ik}g^{jl}h_{kl}$, but is defined as the inverse of $h_{ij}$ and is independent of $g_{ij}$. With such a definition and eq.~(5), we can see that there exist highly non-linear terms composed of $\frac{{\partial}X^{\mu}}{{\partial}{\xi}^{i}}\frac{{\partial}X^{\nu}}{{\partial}{\xi}^{j}}{\eta}_{{\mu}{\nu}}$ in $h^{ij}$.

In this paper we make Hamiltonian analyses for $A_P$, $A^{\rm I}_{W}$, and $A^{\rm II}_{W}$, respectively. Note that the Hamiltonian analysis (including also an auxiliary field) was applied to~\cite{s14} Weyl-invariant strings, D-strings, and to general relativity, and that the applications to the manifestly diffeomorphism invariant actions of the models were made for the first time. The first problem we encounter is how to re-formulate the actions $A_P$, $A^{\rm I}_{W}$, and $A^{\rm II}_{W}$ explicitly in terms of $\frac{{\partial}X^{\mu}}{{\partial}{\xi}^{i}}\frac{{\partial}X^{\nu}}{{\partial}{\xi}^{j}}{\eta}_{{\mu}{\nu}}$, that is, to compute the inverse of $h_{ij}$. This can be resolved by applying an ADM reparametrization~\cite{s15} to the induced metric. Incidentally, similar application to the auxiliary worldvolume metric appeared in refs.~\cite{s16,s11}. After that we calculate canonical momenta. If we follow the usual Hamiltonian procedure, the second problem might be substitution of velocities by the momenta. However, because of the high non-linearity mentioned above to fulfill such a substitution is almost impossible. The same case has occurred~\cite{s17} when dealing with $S^{{\rm II}}_{W}$ whose non-linearity, less complicated than ours, is  just the expedient of $\frac{{\partial}X^{\mu}}{{\partial}{\xi}^{i}}\frac{{\partial}X^{\nu}}{{\partial}{\xi}^{j}}{\eta}_{{\mu}{\nu}}$ to the power $(p+1)/2$. In ref.~\cite{s17} some modification to the usual Hamiltonian procedure is suggested in order to circumvent the difficult task of replacing velocities by momenta in Hamiltonian densities. We find that this  modification is suitable to be applied in a wider region of including our new actions. In the next section we make a brief review of the modified Hamiltonian procedure. Following this procedure, we then canonically analyse $A_P$, $A^{\rm I}_{W}$, and $A^{\rm II}_{W}$ in sections 3, 4, and 5, respectively. At last, in section 6 we will make a summary and compare our results with that of the first set composed of known dual $p$-brane actions.

\section{Brief review of modified Hamiltonian procedure}
In order for this paper to be self-contained we simply repeat the main context of the modified Hamiltonian procedure~\cite{s17}, but use our notation for the sake of consistency in the paper as a whole.

Consider a general system described by the action
\begin{equation}
S=\int d^{p+1}{\xi}\;{\cal L}({\Psi}^A,{\partial}_{i}{\Psi}^A),
\end{equation}
where ${\Psi}^A$ represent all field variables with index $A$ taking values of number of the variables, and ${\partial}_{i}{\Psi}^A$ stand for $\frac{{\partial}{\Psi}^A}{{\partial}{\xi}^{i}}, i=0,1,\cdots,p$. The canonical momenta conjugate to ${\Psi}^A$ are defined by
\begin{equation}
P_{A} \equiv \frac{{\partial}{\cal L}}{{\partial}{\dot{\Psi}}^A}\;,
\end{equation}
where ${\dot{\Psi}}^A \equiv {\partial}_{0}{\Psi}^A$. If the Hessian matrix is singular, there are, say, $R$ primary constraints
\begin{equation}
{\Omega}_M={\Omega}_M({\Psi}^A,P_{A}), \hspace{5mm}M=1,\cdots,R,
\end{equation}
which come from the definition eq.~(7). The primary Hamiltonian density that generates evolution of the system takes the form
\begin{equation}
{\cal H}=P_{A}{\dot{\Psi}}^A-{\cal L}+{\lambda}^M{\Omega}_M,
\end{equation}
where ${\lambda}^M$ are auxiliary Lagrange multipliers introduced.

The usual procedure at this point is trying to replace velocities ${\dot{\Psi}}^A$ by momenta $P_{A}$. Instead,  one computes the general variation of ${\cal H}$
\begin{equation}
\delta {\cal H}=\delta P_{A}{\dot{\Psi}}^A+P_{A}\delta{\dot{\Psi}}^A+\delta{\lambda}^M{\Omega}_M+{\lambda}^M\delta{\Omega}_M-\delta{\cal L}.
\end{equation}
In terms of the expressions
\begin{eqnarray}
\delta{\Omega}_M&=&\frac{{\partial}{\Omega}_M}{{\partial}{\Psi}^A}{\delta}{\Psi}^A+\frac{{\partial}{\Omega}_M}{{\partial}P_A}\delta{P}_A,\nonumber \\
{\delta}{\cal L}&=&\left(\frac{{\partial}{\cal L}}{{\partial}{\Psi}^A}-{\partial}_a\frac{{\partial}{\cal L}}{{\partial}({\partial}_a{\Psi}^A)}\right){\delta}{\Psi}^A+\frac{{\partial}{\cal L}}{{\partial}{\dot{\Psi}}^A}{\delta}{\dot{\Psi}}^A,
\end{eqnarray}
one thus arrives at the formulation of the general variation
\begin{eqnarray}
\delta {\cal H}&=&\left({P}_A-\frac{{\partial}{\cal L}}{{\partial}{\dot{\Psi}}^A}\right){\delta}{\dot{\Psi}}^A +\left({\dot{\Psi}}^A+{\lambda}^M\frac{{\partial}{\Omega}_M}{{\partial}P_A}\right){\delta}{P}_A \nonumber \\
& &+\left({\lambda}^M\frac{{\partial}{\Omega}_M}{{\partial}{\Psi}^A}-\frac{{\partial}{\cal L}}{{\partial}{\Psi}^A}
+{\partial}_a\frac{{\partial}{\cal L}}{{\partial}({\partial}_a{\Psi}^A)}\right){\delta}{\Psi}^A+{\Omega}_M{\delta}{\lambda}^M,
\end{eqnarray}
where indices $a,b,\cdots,f$ take values $1,2,\cdots,p$ here and in the following sections. From the above formulation together with eq.~(7) one can obtain the generalized Hamiltonian equations
\begin{eqnarray}
& &\frac{{\partial}{\cal H}}{{\partial}{\dot{\Psi}}^A}={P}_A-\frac{{\partial}{\cal L}}{{\partial}{\dot{\Psi}}^A}=0, \nonumber \\
& &\frac{{\partial}{\cal H}}{{\partial}{P}^A}={\dot{\Psi}}^A+{\lambda}^M\frac{{\partial}{\Omega}_M}{{\partial}P_A}, \nonumber \\
& &\frac{{\partial}{\cal H}}{{\partial}{\Psi}^A}={\lambda}^M\frac{{\partial}{\Omega}_M}{{\partial}{\Psi}^A}-\frac{{\partial}{\cal L}}{{\partial}{\Psi}^A}
+{\partial}_a\frac{{\partial}{\cal L}}{{\partial}({\partial}_a{\Psi}^A)}, \nonumber \\
& &\frac{{\partial}{\cal H}}{{\partial}{\lambda}^M}={\Omega}_M.
\end{eqnarray}

The next step in the Hamiltonian procedure is to impose the time conservation of the primary constraints ${\Omega}_M$. This can also be done without replacing velocities by momenta. The time evolution of constraints takes the form
\begin{equation}
{\dot{\Omega}}_M({\xi}^0,{\mbox{\boldmath ${\xi}$}})=\int d^p {\mbox{\boldmath ${\xi}$}}^{\prime}\{{\Omega}_M({\xi}^0,{\mbox{\boldmath ${\xi}$}}),{\cal H}({\xi}^0,{\mbox{\boldmath ${\xi}$}}^{\prime})\} \approx 0,
\end{equation}
where ``$\approx$'' means Dirac's weak equality~\cite{s18}. By using the derivative properties of Poisson brackets
\begin{equation}
\{{\Omega}_M,{\cal L}\}=\{{\Omega}_M,{\Psi}^A\}\frac{{\partial}{\cal L}}{{\partial}{\Psi}^A}+\{{\Omega}_M,{\dot{\Psi}}^A\}\frac{{\partial}{\cal L}}{{\partial}{\dot{\Psi}}^A}+\{{\Omega}_M,{\partial}_a{\Psi}^A\}\frac{{\partial}{\cal L}}{{\partial}({\partial}_a{\Psi}^A)},
\end{equation}
and eq.~(9) one calculates the Poissson bracket between ${\Omega}_M$ and ${\cal H}$
\begin{eqnarray}
\{{\Omega}_M,{\cal H}\}&=&\{{\Omega}_M,{\dot{\Psi}}^A\}\left({P}_A-\frac{{\partial}{\cal L}}{{\partial}{\dot{\Psi}}^A}\right)+\{{\Omega}_M,{P}_A\}{\dot{\Psi}}^A-\{{\Omega}_M,{\Psi}^A\}\frac{{\partial}{\cal L}}{{\partial}{\Psi}^A}\nonumber \\
& &-\{{\Omega}_M,{\partial}_a{\Psi}^A\}\frac{{\partial}{\cal L}}{{\partial}({\partial}_a{\Psi}^A)}+{\lambda}^N\{{\Omega}_M,{\Omega}_N\}.
\end{eqnarray}
Although $\{{\Omega}_M,{\dot{\Psi}}^A\}$ is unknown because the replacement of ${\dot{\Psi}}^A$ by ${P}_A$ is quite hard to be realized, the first term of eq.~(16) vanishes in accordance with eq.~(7). Consequently, this difficult task has been circumvented. Further constraints may be deduced by imposing the weak equality~(14) whose integrand, {\em i.e.} the Poissson bracket of ${\Omega}_M$ and ${\cal H}$,  is now available
\begin{equation}
\{{\Omega}_M,{\cal H}\}=\{{\Omega}_M,{P}_A\}{\dot{\Psi}}^A-\{{\Omega}_M,{\Psi}^A\}\frac{{\partial}{\cal L}}{{\partial}{\Psi}^A}
-\{{\Omega}_M,{\partial}_a{\Psi}^A\}\frac{{\partial}{\cal L}}{{\partial}({\partial}_a{\Psi}^A)}+{\lambda}^N\{{\Omega}_M,{\Omega}_N\}.
\end{equation}
Eq.~(14), together with eq.~(17), is either identically satisfied, or it leads to determination of the Lagranger multipliers ${\lambda}^M$, or it implies existence of new constraints. In the last case the procedure has to be continued in a similar way until all the descendant (secondary, tertiary, {\em etc.}) constraints are finally derived. Note that it is only at the end of the process that one tries the substitution of velocities by momenta.

\section{Hamiltonian analysis of $A_P$}
With the modified Hamiltonian procedure in mind we now begin to do the Hamiltonian analysis for $A_P$. At first, we have to rewrite eq.~(1) explicitly in the expression composed of ${\partial}_0X^{\mu}{\partial}_iX_{\mu}$ in order to calculate canonical momenta conjugate to $X^{\mu}$, that is, we have to compute $h^{ij}$ at first. To this end, we make the ADM reparametrization~\cite{s15} of the induced metric $h_{ij}$ in terms of a shift-vector $N^a$, a lapse function $N$ and a $p$-metric ${\gamma}_{ab}$, all of which depend on ${\xi}^i$:
\begin{eqnarray}
& &h_{00}=-N^2+{\gamma}_{ab}N^aN^b, \nonumber\\
& &h_{0a}=h_{a0}={\gamma}_{ab}N^b, \nonumber\\
& &h_{ab}={\gamma}_{ab},
\end{eqnarray}
and
\begin{eqnarray}
& &h^{00}=-N^{-2}, \nonumber\\
& &h^{0a}=h^{a0}=N^aN^{-2}, \nonumber\\
& &h^{ab}={\gamma}^{ab}-N^aN^bN^{-2},
\end{eqnarray}
where ${\gamma}^{ab}$ is the inverse of ${\gamma}_{ab}$. Substituting the definition of the induced metric eq.~(5) into eq.~(18), we solve $N^a$ and $N$ in terms of ${\partial}_0X^{\mu}{\partial}_iX_{\mu}$:
\begin{eqnarray}
N^a&=&{\gamma}^{ab}({\dot X}{\partial}_{b}X), \nonumber\\
N^2&=&-{\dot X}^2+{\gamma}^{ab}({\dot X}{\partial}_{a}X)({\dot X}{\partial}_{b}X),
\end{eqnarray}
where the sum of Lorentzian indices has been suppressed. Using eqs.~(19) and~(20) we therefore obtain the inverse of $h_{ij}$ as follows:
\begin{eqnarray}
& &h^{00}=-\frac{1}{-{\dot X}^2+{\gamma}^{ab}({\dot X}{\partial}_{a}X)({\dot X}{\partial}_{b}X)}, \nonumber\\
& &h^{0a}=h^{a0}=\frac{{\gamma}^{ab}({\dot X}{\partial}_{b}X)}{-{\dot X}^2+{\gamma}^{cd}({\dot X}{\partial}_{c}X)({\dot X}{\partial}_{d}X)}, \nonumber\\
& &h^{ab}={\gamma}^{ab}-\frac{{\gamma}^{ac}{\gamma}^{bd}({\dot X}{\partial}_{c}X)({\dot X}{\partial}_{d}X)}{-{\dot X}^2+{\gamma}^{ef}({\dot X}{\partial}_{e}X)({\dot X}{\partial}_{f}X)}.
\end{eqnarray}
Note that ${\gamma}^{ab}$ is independent of ${\partial}_0X^{\mu}$, and that $h^{ab} \neq {\gamma}^{ab}$ though $h_{ab}={\gamma}_{ab}$ because $h_{ij}$ is a $(p+1)$-metric while ${\gamma}_{ab}$ a $p$-metric. With the help of eq.~(21) we re-formulate eq.~(1) to be
\begin{eqnarray}
{A}_{P}&=&-\frac{T}{2}\int d^{p+1}{\xi}\sqrt{-g}\Bigg[\frac{g_{00}-2g_{0a}{\gamma}^{ab}({\dot X}{\partial}_{b}X)+g_{ab}{\gamma}^{ac}{\gamma}^{bd}({\dot X}{\partial}_{c}X)({\dot X}{\partial}_{d}X)}{-{\dot X}^2+{\gamma}^{ef}({\dot X}{\partial}_{e}X)({\dot X}{\partial}_{f}X)}\nonumber\\
& &-g_{ab}{\gamma}^{ab}+(p+3)\Bigg].
\end{eqnarray}

Eqs.~(7) and~(22) give the canonical momenta conjugate to $g_{ij}$,
\begin{equation}
{\Pi}^{ij} \approx 0,
\end{equation}
which mean $\frac{1}{2}(p+1)(p+2)$ primary constraints, and that conjugate to $X^{\mu}$,
\begin{eqnarray}
P_{\mu}&=&{T}\sqrt{-g}\left[-{\dot X}^2+{\gamma}^{ab}({\dot X}{\partial}_{a}X)({\dot X}{\partial}_{b}X)\right]^{-2}\nonumber\\
& \times&   \Bigg\{\left[-{\dot X}_{\mu}+{\gamma}^{ab}{\partial}_{a}X_{\mu}({\dot X}{\partial}_{b}X)\right]\left[g_{00}-2g_{0a}{\gamma}^{ab}({\dot X}{\partial}_{b}X)+g_{ab}{\gamma}^{ac}{\gamma}^{bd}({\dot X}{\partial}_{c}X)({\dot X}{\partial}_{d}X)\right]\nonumber\\
& & - \left[-g_{0a}{\gamma}^{ab}{\partial}_{b}X_{\mu}+g_{ab}{\gamma}^{ac}{\gamma}^{bd}{\partial}_{c}X_{\mu}({\dot X}{\partial}_{d}X)\right]\left[-{\dot X}^2+{\gamma}^{ab}({\dot X}{\partial}_{a}X)({\dot X}{\partial}_{b}X)\right]\Bigg\},
\end{eqnarray}
respectively. It is obvious from eq.~(24) that to solve ${\dot X}^{\mu}$ in terms of $P_{\mu}$ is quite difficult and almost impossible. Following the procedure reviewed briefly in the above section, we now compute the time evolution of the primary constraints~(23) by using eqs.~(14),~(17), and~(22)
\begin{equation}
{\dot{\Pi}}^{ij}=-\frac{T}{2}\sqrt{-g}\left[\frac{1}{2}g^{ij}\left(-g_{kl}h^{kl}+p+3\right)-h^{ij}\right] \approx 0,
\end{equation}
from which we obtain the solution called sometimes an embedding relation
\begin{equation}
g_{ij}=h_{ij}.
\end{equation}
When substituting eq.~(26) into eq.~(1), $A_{P}$ turns to $S_{NG}$, which shows the classical equivalence between the two action as was discussed in detail in ref.~\cite{s2}. Here, however, in terms of $00$ and $0a$ components of eq.~(26) we eliminate velocities in eq.~(24) and derive the following relations that in fact mean $p+1$ constraints associated with the invariance of $A_{P}$ under $p+1$ reparametrizations
\begin{eqnarray}
\left[g^{00}\left(g_{00}-g_{0a}g_{0b}{\gamma}^{ab}\right)+\frac{g_{0a}{\gamma}^{ab}(P{\partial}_{b}X)}{{T}\sqrt{-g}}\right]^2&=&-\frac{1}{T^2gg^{00}}\left[P^2-{\gamma}^{ab}(P{\partial}_{a}X)(P{\partial}_{b}X)\right],\nonumber\\
g_{0a}-g_{ab}g_{0c}{\gamma}^{bc}&=&-\frac{1}{g^{00}}\:\frac{(P{\partial}_{a}X)}{{T}\sqrt{-g}}.
\end{eqnarray}

Before looking for further constraints let us classify these constraints into first- and second-classes. The first-class ones consist of $00$ and $0a$ components of eq.~(23)
\begin{eqnarray}
{\Pi}^{00} &\approx & 0,\nonumber\\
{\Pi}^{0a} &\approx &0,
\end{eqnarray}
and eq.~(27) which can be reduced to the usual forms by using $ab$ components of eq.~(26)
\begin{eqnarray}
P^2+T^2{\gamma}&\approx &0,\nonumber\\
(P{\partial}_{a}X)&\approx &0,
\end{eqnarray}
where ${\gamma}\equiv {\rm det}({\gamma}_{ab})$. The second-class constraints contain $ab$ components of eq.~(23) and of eq.~(26)
\begin{eqnarray}
{\Pi}^{ab} &\approx & 0,\nonumber\\
g_{ab}-{\gamma}_{ab}&\approx & 0.
\end{eqnarray}
The number of independent degrees of freedom in phase space is given by the total number of degrees of freedom $2D+(p+1)(p+2)$ subtracted by twice of the number of first-class constraints $4(p+1)$ and by the number of second-class ones $p(p+1)$, that is, $2D-2(p+1)$. This is the correct number, which shows that no more constraints exist. As a result, we find all the constraints by following the procedure introduced in section 2. In particular, the difficult task of replacing velocities by momenta in Hamiltonian densities has been circumvented. The remaining procedure can be continued as dealt with, for instance, in refs.~\cite{s16,s11} but is omitted in this and the next two sections as in ref.~\cite{s17}.

\section{Hamiltonian analysis of $A^{\rm I}_{W}$}
Substituting the ADM transformation~(21) into eq.~(2), we rewrite $A^{\rm I}_{W}$ to be
\begin{eqnarray}
A^{{\rm I}}_{W}&=&-\frac{T}{2}\int d^{p+1}{\xi}\sqrt{-g}\Bigg[\frac{g_{00}-2g_{0a}{\gamma}^{ab}({\dot X}{\partial}_{b}X)+g_{ab}{\gamma}^{ac}{\gamma}^{bd}({\dot X}{\partial}_{c}X)({\dot X}{\partial}_{d}X)}{-{\dot X}^2+{\gamma}^{ef}({\dot X}{\partial}_{e}X)({\dot X}{\partial}_{f}X)}\:{\Phi}^{(p+3)/2}\nonumber\\
& &-g_{ab}{\gamma}^{ab}{\Phi}^{(p+3)/2}+(p+3){\Phi}^{(p+1)/2}\Bigg].
\end{eqnarray}
The canonical momenta conjugate to $g_{ij}$ take the same form as eq.~(23), and the momenta conjugate to ${\Phi}$ and to $X^{\mu}$ are given respectively by
\begin{equation}
{\Pi}_{\Phi} \approx 0,
\end{equation}
which means a primary constraint, and
\begin{eqnarray}
P_{\mu}&=&{T}\sqrt{-g}\:{\Phi}^{(p+3)/2}\left[-{\dot X}^2+{\gamma}^{ab}({\dot X}{\partial}_{a}X)({\dot X}{\partial}_{b}X)\right]^{-2}\nonumber\\
& \times&   \Bigg\{\left[-{\dot X}_{\mu}+{\gamma}^{ab}{\partial}_{a}X_{\mu}({\dot X}{\partial}_{b}X)\right]\left[g_{00}-2g_{0a}{\gamma}^{ab}({\dot X}{\partial}_{b}X)+g_{ab}{\gamma}^{ac}{\gamma}^{bd}({\dot X}{\partial}_{c}X)({\dot X}{\partial}_{d}X)\right]\nonumber\\
& & - \left[-g_{0a}{\gamma}^{ab}{\partial}_{b}X_{\mu}+g_{ab}{\gamma}^{ac}{\gamma}^{bd}{\partial}_{c}X_{\mu}({\dot X}{\partial}_{d}X)\right]\left[-{\dot X}^2+{\gamma}^{ab}({\dot X}{\partial}_{a}X)({\dot X}{\partial}_{b}X)\right]\Bigg\}.
\end{eqnarray}
According to the modification of the Hamiltonian procedure introduced in section 2, we calculate the time preservation of the primary constraints eqs.~(23) and~(32)
\begin{eqnarray}
{\dot{\Pi}}^{ij}&=&-\frac{T}{2}\sqrt{-g}\:{\Phi}^{(p+1)/2}\left[\frac{1}{2}g^{ij}\left(-{\Phi}g_{kl}h^{kl}+p+3\right)-{\Phi}h^{ij}\right] \approx 0,\\
{\dot{\Pi}}_{{\Phi}}&=&-\frac{T(p+3)}{4}\sqrt{-g}\:{\Phi}^{(p-1)/2}\left(-{\Phi}g_{ij}h^{ij}+p+1\right) \approx 0.
\end{eqnarray}
From eq.~(35) we have
\begin{equation}
{\Phi}^{-1}=\frac{1}{p+1}\:g_{ij}h^{ij},
\end{equation}
and substituting eq.~(36) into eq.~(34) gives the embedding relation
\begin{equation}
g_{ij}=f(\xi)h_{ij},
\end{equation}
where $f(\xi)$ is an arbitrary function of ${\xi}^i$. The appearance  of $f(\xi)$ in eq.~(37) shows that $A^{{\rm I}}_{W}$ possesses the Weyl symmetry as we mentioned before. When substituting eqs.~(36) and~(37) into eq.~(2), $A^{\rm I}_{W}$ reduces to $S_{NG}$. This expresses the classical equivalence between the two actions. Similar to the treatment in the above section, we eliminate velocities in eq.~(33) by using $00$ and $0a$ components of eq.~(37) and then obtain the following $p+1$ constraints that associate with the invariance of the action $A^{\rm I}_{W}$ under $p+1$ reparametrizations
\begin{eqnarray*}
& &\left[g^{00}\left(g_{00}-f^{-1}g_{0a}g_{0b}{\gamma}^{ab}\right)+f^{-2}{\Phi}^{-(p+3)/2}\:\frac{g_{0a}{\gamma}^{ab}(P{\partial}_{b}X)}{{T}\sqrt{-g}}\right]^2  \\
&=&-f^{-3}{\Phi}^{-(p+3)}\:\frac{1}{T^2gg^{00}}\left[P^2-{\gamma}^{ab}(P{\partial}_{a}X)(P{\partial}_{b}X)\right],
\end{eqnarray*}
\begin{equation}
g_{0a}-f^{-1}g_{ab}g_{0c}{\gamma}^{bc}=-f^{-1}{\Phi}^{-(p+3)/2}\:\frac{1}{g^{00}}\:\frac{(P{\partial}_{a}X)}{{T}\sqrt{-g}}.
\end{equation}
Although eq.~(38) looks more complicated than eq.~(27), it reduces exactly to eq.~(29) after eq.~(36) and $ab$ components of eq.~(37) are considered. Therefore, the system described by $A^{\rm I}_{W}$ contains the same set of first-class constraints as expressed by eqs.~(28) and~(29). As to second-class,  from eqs.~(23),~(32),~(36), and~(37) we find that the corresponding set consists of the following $p(p+1)+2$ constraints
\begin{eqnarray}
{\Pi}^{ab} \approx 0, & &\hspace{5mm}g_{ab}-f{\gamma}_{ab} \approx 0, \\
{\Pi}_{\Phi} \approx 0, & &\hspace{5mm}{\Phi}-f^{-1}\approx 0.
\end{eqnarray}
The number of independent degrees of freedom in phase space is $[2D+(p+1)(p+2)+2]-4(p+1)-[p(p+1)+2]=2D-2(p+1)$ as expected.

\section{Hamiltonian analysis of $A^{\rm II}_{W}$}
Substituting the ADM transformation~(21) into eq.~(3), we rewrite $A^{\rm II}_{W}$ to be
\begin{eqnarray}
A^{{\rm II}}_{W}&=&-T\int d^{p+1}{\xi}\sqrt{-g}\Bigg[\frac{1}{p+1}\Bigg(-\frac{g_{00}-2g_{0a}{\gamma}^{ab}({\dot X}{\partial}_{b}X)+g_{ab}{\gamma}^{ac}{\gamma}^{bd}({\dot X}{\partial}_{c}X)({\dot X}{\partial}_{d}X)}{-{\dot X}^2+{\gamma}^{ef}({\dot X}{\partial}_{e}X)({\dot X}{\partial}_{f}X)}\nonumber\\
& &+g_{ab}{\gamma}^{ab}\Bigg)\Bigg]^{-(p+1)/2}.
\end{eqnarray}
The canonical momenta conjugate to $g_{ij}$ are given by eq.~(23) as before, and the momenta conjugate to $X^{\mu}$, however, take the form
\begin{eqnarray}
P_{\mu}&=&{T}\sqrt{-g}\:\left(\frac{1}{p+1}g_{ij}
h^{ij}\right)^{-(p+3)/2}\left[-{\dot X}^2+{\gamma}^{ab}({\dot X}{\partial}_{a}X)({\dot X}{\partial}_{b}X)\right]^{-2}\nonumber\\
& \times&   \Bigg\{\left[-{\dot X}_{\mu}+{\gamma}^{ab}{\partial}_{a}X_{\mu}({\dot X}{\partial}_{b}X)\right]\left[g_{00}-2g_{0a}{\gamma}^{ab}({\dot X}{\partial}_{b}X)+g_{ab}{\gamma}^{ac}{\gamma}^{bd}({\dot X}{\partial}_{c}X)({\dot X}{\partial}_{d}X)\right]\nonumber\\
& & - \left[-g_{0a}{\gamma}^{ab}{\partial}_{b}X_{\mu}+g_{ab}{\gamma}^{ac}{\gamma}^{bd}{\partial}_{c}X_{\mu}({\dot X}{\partial}_{d}X)\right]\left[-{\dot X}^2+{\gamma}^{ab}({\dot X}{\partial}_{a}X)({\dot X}{\partial}_{b}X)\right]\Bigg\}.
\end{eqnarray}
With the modified Hamiltonian procedure in mind we compute the time evolution of the primary constraints~(23) by using eqs.~(14),~(17), and~(41)
\begin{equation}
{\dot{\Pi}}^{ij}=-\frac{T}{2}\sqrt{-g}\:\left(\frac{1}{p+1}g_{kl}
h^{kl}\right)^{-(p+3)/2}\left(\frac{1}{p+1}g^{ij}g_{kl}h^{kl}-h^{ij}\right) \approx 0,
\end{equation}
which leads to the same solution as eq.~(37). Similarly, the existence of an arbitrary function reveals the conformal invariance of $A^{\rm II}_{W}$.
Substituting eq.~(37) into eq.~(3) makes $A^{\rm II}_{W}$ become $S_{NG}$, which gives the classical equivalence between the two actions.
As treated in the above section we derive $p+1$ constraints from eq.~(42) by using eq.~(37)
\begin{eqnarray*}
& &\left[g^{00}\left(g_{00}-f^{-1}g_{0a}g_{0b}{\gamma}^{ab}\right)+f^{(p-1)/2}\:\frac{g_{0a}{\gamma}^{ab}(P{\partial}_{b}X)}{{T}\sqrt{-g}}\right]^2\\
&=&-\frac{f^p}{T^2gg^{00}}\left[P^2-{\gamma}^{ab}(P{\partial}_{a}X)(P{\partial}_{b}X)\right],
\end{eqnarray*}
\begin{equation}
g_{0a}-f^{-1}g_{ab}g_{0c}{\gamma}^{bc}=-f^{(p+1)/2}\:\frac{1}{g^{00}}\:\frac{(P{\partial}_{a}X)}{{T}\sqrt{-g}},
\end{equation}
which are related with the invariance of the action $A^{\rm II}_{W}$ under $p+1$ reparametrizations. Again using $ab$ components of eq.~(37) we can simplify eq.~(44) to be eq.~(29). That is, this system has the same set of first-class constraints as that of $A_P$ and of $A^{\rm I}_{W}$. Its second-class set is nevertheless given by eq.~(39). The independent number of degrees of freedom in phase space is, of course, $2D-2(p+1)$ as we have known.

\section{Conclusion}
We have made Hamiltonian analyses to $A_P$, $A^{\rm I}_{W}$, and $A^{\rm II}_{W}$ which, together with $S_{NG}$, constitute the new set of dual actions of bosonic $p$-branes. Because of the special definition of the contravariant components of the induced metric, we have utilized the ADM reparametrization to rewrite the actions explicitly in terms of ${\partial}_0X^{\mu}{\partial}_iX_{\mu}$, and therefore we have been able to compute canonical momenta conjugate to $X^{\mu}$ defined as usual. What caused by such a definition of $h^{ij}$ is that the actions contain highly non-linear terms composed of ${\partial}_0X^{\mu}{\partial}_iX_{\mu}$, which makes it almost impossible to replace velocities by momenta in Hamiltonian densities when following the usual canonical procedure. Fortunately, it has been discovered that such a replacement is not mandatory when canonically analysing the conformally invariant $p$-brane action $S^{\rm II}_{W}$, and some modification to circumvent this difficult task has been suggested. What we have developed in this paper is that this modification of the Hamiltonian analysis has a wider application that includes our newly constructed $p$-brane actions that contain higher non-linearity than that of $S^{\rm II}_{W}$. Quite interesting is that the three new actions have the same first-class set of constraints, while their second-class sets are different from each other. The similar case also happens in the known $p$-brane actions $S_{P}$ and $S^{{\rm II}}_{W}$. Further studies, such as various possible supersymmetric extensions of the new actions and their canonical quantization, are now under consideration.

\vspace{10mm}
\noindent
{\bf Acknowledgments}

The author would like to thank Nobuyoshi Ohta for helpful discussions and comments, and to acknowledge the support through him from the Grant-in-Aid for Scientific Research. He is also supported in part by
the National Natural Science Foundation of China under grant No.10275052 and by Nankai University under grant No.J02013.


\newpage
\baselineskip 20pt
\begin{thebibliography}{s3}
\bibitem{s1}For a review, see, M.J. Duff, {\em Supermembranes}, hep-th/9611203.
\bibitem{s2}Y.-G. Miao and N. Ohta, {\em J. of High Energy Phys.} {\bf 04} (2003) 010 [hep-th/0301233].
\bibitem{s3}S. Deser and R. Jackiw, {\em Phys. Lett.} {\bf B 139} (1984) 371;
S.E. Hjelmeland and U. Lindstr\"om, {\em Duality for the nonspecialist}, hep-th/9705122.
\bibitem{s4}Y. Nambu, {\em Lectures at the Copenhagen conference}, 1970.
\bibitem{s5}T. Goto, {\em Prog. Theor. Phys.} {\bf 46} (1971) 1560.
\bibitem{s6}P.S. Howe and R.W. Tucker, {\em J. Phys.} {\bf A 10} (1977) L155;
{\em J. Math. Phys.} {\bf 19} (1978) 981.
\bibitem{s7}L. Brink, P. Di Vecchia, and P. Howe, {\em Phys. Lett.}
{\bf B 65} (1976) 471.
\bibitem{s8}S. Deser and B. Zumino, {\em Phys. Lett.} {\bf B 65} (1976) 369.
\bibitem{s9}A.M. Polyakov, {\em Phys. Lett.} {\bf B 103} (1981) 207.
\bibitem{s10}U. Lindstr\"om, {\em Int. J. Mod. Phys.} {\bf A 3} (1988) 2401.
\bibitem{s11}J.A. Garcia, R. Linares, and J.D. Vergara, {\em Phys. Lett.}
{\bf B 503} (2001) 154 [hep-th/0011085].
\bibitem{s12}J.A. Nieto, {\em Mod. Phys. Lett.} {\bf A 16} (2001) 2567
[hep-th/0110227].
\bibitem{s13}S. Deser, M.J. Duff, and C.J. Isham, {\em Nucl. Phys.} {\bf B 114}
(1976) 29.
\bibitem{s14}U. Lindstr\"om and H.G. Svendsen, {\em Int. J. Mod. Phys.} {\bf A 16} (2001) 1347 [hep-th/0007101].
\bibitem{s15}See, for example, C.W. Misner, K.S. Thorne, and J.A. Wheeler, {\em Gravitation}, W.M. Freeman, San Francisco 1970.
\bibitem{s16}M.J. Duff, T. Inami, C.N. Pope, E. Sezgin, and K.S. Stelle, {\em Nucl. Phys.} {\bf B 297}
(1988) 515.
\bibitem{s17}C. Alvear, R. Amorim, and J. Barcelos-Neto, {\em Phys. Lett.}
{\bf B 273} (1991) 415.
\bibitem{s18}P.A.M. Dirac, {\em Lectures on Quantum Mechanics}, Yeshiva University, New York 1964.
\end{thebibliography}
\end{document}